\documentclass[12pt]{article}
\usepackage{amsmath,amssymb,amsthm,latexsym}
\usepackage{wasysym,upgreek,mathrsfs}
\usepackage[english]{babel}
\usepackage{graphicx,color}
\usepackage{amscd}

\usepackage[pdftex]{hyperref}
\newtheorem{lemma}{Lemma}[section]
\newtheorem{corollary}{Corollary}[section]

\newtheorem{definition}{Definition}[section]
\newtheorem{theorem}{Theorem}[section]
\newcommand{\be}{\begin{equation}}
\newcommand{\bee}{\begin{equation}}
\newcommand{\ee}{\end{equation}}

\newcommand{\beqa}{\begin{eqnarray}}
\newcommand{\eeqa}{\end{eqnarray}}

\newcommand{\bea}{\begin{eqnarray}}
\newcommand{\eea}{\end{eqnarray}}

\newcommand{\prf}{\noindent {\rm\bf  Proof:\;\; }}

\newcommand{\cA}{{{\cal A}}}

\newcommand{\cF}{{\cal{F}}}
\newcommand{\cS}{{\cal{S}}}

\begin{document}

 \title{Generalized Constructive {Tree} Weights}
\author{Vincent Rivasseau$^{a,b}$,\  Adrian Tanasa$^{c,d}$\\
\bigskip
\\
E-mail:\\
 vincent.rivasseau@th.u-psud.fr, \\
adrian.tanasa@ens-lyon.org}

\maketitle
\begin{abstract}
The Loop Vertex Expansion (LVE) is a quantum field theory (QFT) method which 
explictly computes the Borel sum of Feynman perturbation series. 
This LVE relies in a crucial way on symmetric tree weights which define a measure on the set of spanning trees of 
any connected graph. In this paper we generalize this method by
defining new tree weights. They depend on the choice of a partition of a set of vertices of the graph, and when the partition is non-trivial,
they are no longer symmetric under permutation of vertices.
Nevertheless we prove they have the required positivity property to lead to a convergent LVE; in fact we formulate
this positivity property precisely for the first time. 
Our generalized tree weights are inspired by the Brydges-Battle{-}Federbush work on cluster expansions
and could be particularly suited to the computation of connected functions in QFT.
Several concrete examples are explicitly given.
\end{abstract}

\begin{flushright}
\end{flushright}
\medskip

\noindent  Key words: Trees, Feynman graphs, Combinatorics, Constructive {quantum} field theory.

\medskip

\section{Introduction}

The fundamental step in quantum field theory (QFT) is to compute the logarithm of a functional integral\footnote{The main 
feature of QFT is the renormalization group, which is made of a sequence of such fundamental steps, one for each \emph{scale}.}.
{This comes from a fundamental theorem of enumerative combinatorics, stating the logarithm counts the connected objects.}
The main advantage of the perturbative expansion of a QFT into a sum of Feynman amplitudes is to
perform this computation explicitly: the logarithm of the functional integral 
is simply the same sum of Feynman amplitudes restricted to \emph{connected} graphs. The main disadvantage 
is that the perturbative series indexed by Feynman graphs typically \emph{diverges}. Constructive theory 
is the right compromise, which allows both to compute logarithms, hence
connected quantities, but through \emph{convergent series}. However it has the reputation to be 
a difficult technical subject.

Perturbative QFT writes quantities of interest (free energies or
connected functions) as sums of amplitudes of connected graphs
\be  S = \sum_{G} A_G . \label{ordinar}
\ee
However such a formula (obtained by expanding in a power series the exponential of the interaction in Feynman functional integral
and then {"}illegally{"} commuting the
power series and the functional integration) is \emph{not} a valid definition since
usually, even with cutoffs, even in zero dimension (!) we have 
\be  \sum_{G} \vert A_G \vert = \infty .
\ee
This divergence, known since \cite{Dys},
is due to the more-than-exponential growth of the number of graphs with many vertices. 
We can say that Feynman graphs {\it  proliferate too fast}. 
More precisely the power series in the coupling constant $\lambda$ corresponding to \eqref{ordinar}
has zero radius of convergence\footnote{This can be proved easily for $\phi^4_d$, the Euclidean Bosonic QFT
with quartic interaction in dimension $d$, with fixed ultraviolet cutoff,
where the series behaves as $ \sum_n  (-\lambda)^n K^n n! $. It is expected to remain true also 
for the renormalized series without cutoff; this has been proved in the  super-renormalizable cases $d=2,3$ \cite{Jaf,CR}).}.
Nevertheless for the stable Bosonic models which have been rigorously 
built by constructive field theory, the constructive answer is always the \emph{Borel sum}
of the perturbative series (see \cite{Riv} and references therein). Hence the perturbative expansion, although divergent, contains all the
information of the theory; but it should be \emph{reshuffled} into a convergent sum\footnote{Although this may sound like a technical question,
convergence is such a critical issue in QFT that new technical tools for its solution might lead,
like Feynman graphs themselves, to new physical insights.}.

The key point for the success of constructive theory is that 
{\it trees} do {\it not} proliferate as fast as graphs, and they are sufficient to show connectivity, 
hence to compute logarithms. This central fact is not emphasized as such in the classical constructive literature \cite{GJ}. It is usually
obscured by the historic tools which constructive theory borrowed from statistical mechanics, namely lattice cluster and Mayer expansions. 

The {loop vertex expansion} (LVE) \cite{R1} is a relatively recent simple constructive technique which precisely reshuffles ordinary perturbative expansion 
into a convergent expansion using \emph{canonical combinatoric tools} rather than non-canonical lattices.
Initially introduced to analyze \emph{matrix} models with quartic interactions, it has been 
extended to many other stable interactions in \cite{RW} and shown compatible with 
direct space decay estimates \cite{MR1} and with renormalization in simple super-renormalizable cases 
\cite{Rivasseau:2011df,ZWGW}.
It has also recently been used in the context of group field theory \cite{MNRS}
and improved \cite{Gurau} to organize the $1/N$
expansion \cite{expansion1,expansion2,expansion3} for random \emph{tensors} models \cite{review,universality,uncoloring},
a promising approach to random geometry and quantum gravity in more
than two dimensions \cite{TT1,TT2}.

The combinatoric core of the LVE has been reformulated in a more transparent way in  \cite{Rivasseau:2013ova}.
The basic idea is to define a set of positive weights $w(G,T)$ which are 
associated to any pair made of a connected graph $G$ and a spanning tree
$T \subset G$. They are normalized so as to form a probability measure on the spanning trees of $G$:
\be  \sum_{T \subset G}  w (G,T) =1  \label{bary0} .
\ee
To compute constructively instead of perturbatively a QFT quantity $S$ one should use 
{equation} \eqref{bary0} 
to introduce a sum over trees for each graph, and then simply
\emph{exchange} the order of summation between graphs and their spanning trees
\be  S = \sum_{G} A_G = \sum_{G }  \sum_{T \subset G}  w(G,T)  A_G =  \sum_{T} A_T, \quad A_T = \sum_{G \supset T}  w(G,T)  A_G .
\label{const}
\ee

The constructive "miracle" is that if one uses the \emph{right graphs} and \emph{right weights},
then for stable Bosonic interactions with cutoffs we get
\be  \sum_{T}  \vert A_T\vert < +\infty ,
\label{const1}
\ee
which means that $S$ is now well defined; furthermore the result is the desired one, namely the Borel sum of the ordinary perturbation expansion.

The three main tasks are then to identify the most general class of \emph{right weights} to use for \eqref{const1} 
to hold, then the most general class of interactions leading to \emph{right graphs} so that \eqref{const1} holds,
and finally to extend the formalism to include \emph{renormalization}, hence to treat QFTs without cutoffs.

The essential discovery of the LVE \cite{R1} is that even in the case of the simplest stable $\phi^4$ quartic interaction
the "right graphs" to use  for  \eqref{const1} to hold are not the ordinary Feynman graphs, 
but the Feynman graphs of the so-called intermediate field representation of the theory.
This representation decomposes ordinary interactions of order higher than three into three-body interactions. 
This result, initially limited to quartic interactions, has been extended to include all even monomials in \cite{RW}. 

Consider from now on a positive interaction such as $\phi^4$ and its intermediate 
field perturbation expansion. Not every probability measure $w(G,T)$ on the spanning trees of the corresponding graphs
leads to a constructive reshuffling, namely one for which \eqref{const1} holds. For instance the trivial, equally-distributed weights 
$1/ \chi (G)$, where $\chi(G)$, the complexity of $G$, is the number of its spanning trees,
form obviously such a probability measure, but there is no reason
to think they lead to a constructive reshuffling.

The LVE weights used in \cite{R1} are defined in terms of 
the Taylor forest formula of \cite{BK,AR1}, because this formula has a particular positivity property.
This is why they lead to a convergent LVE, for which \eqref{const1} holds. 

These LVE weights are fully symmetric under action of the permutation group
of the vertices of the graph, hence from now on we call denote them $w_s (G,T)$. 
In \cite{Rivasseau:2013ova} these symmetric weights were identified 
with the percentage of Hepp sectors \cite{Hepp} in which the tree $T$ is leading, in the sense 
of Kruskal "greedy algorithm" \cite{kruskal} {(see below)}. 

However, historically, non-symmetric forest formulas for constructive cluster expansions were discovered 
before the symmetric Taylor forest formula (see \cite{BF1}{,} \cite{BF2} {or} \cite{BF3}).
It is therefore clear that the symmetric weights are not the only ones with constructive positivity. 
In this paper we explore this issue in detail. First we define precisely this property of
constructive positivity. Then we identify a very general class of weights, associated
to any non-trivial partition of the vertices of a graph, which have this constructive 
property\footnote{This class may even be the most general one with constructive
positivity but we do not know how to prove or disprove this last point.}.
This is the main result of this paper.

Examples are then given: in particular weights for {\it rooted} and {\it multi-rooted graphs} (which just correspond to particular cases of vertex partitions, where one considers one or more singletons and the rest of vertices are grouped into a single remaining block).
are defined and compared to the symmetric weights. Although
symmetric weights look the most natural for the constructive expansion of quantities such as the free energy of a 
QFT, non-symmetric rooted
weights could be also useful, in particular to compute Schwinger functions with external arguments.
Indeed the latter correspond to particular marked vertices bearing cilia in the LVE \cite{Gurau},
and there is no reason to think that the optimal LVE in that case should be so symmetric as to mix 
these particular external marked vertices with the ordinary internal vertices. 


\section{Prerequisites}

Consider a fixed set $V$ of vertices. We associate weights $w(G,T)$ 
(also called \emph{amplitudes} in the QFT context) 
to any pair $(G,T)$, where $G$ is a labeled connected graph $G$ with vertex set $V$ and edge set $E$,
and $T$ is a spanning tree of $G$ 
(where by spanning tree we mean an acyclic maximal subset of $E$, hence of cardinality $\vert V \vert -1$). 
Self-loops ({\it a. k. a.} tadpoles in the QFT language) and multiple edges in $G$ are allowed, since they are a fundamental feature of QFT. 
From now on we omit the word "spanning", since throughout this paper the trees considered are always spanning
for a related graph $G \supset T$.

We write $\sum_{T \subset G}  w(G,T)$ to indicate summation over the finite family of trees of a fixed $G$
of such weights $w(G,T)$.
We can also consider trees $T$ as  particular labeled connected graph themselves with vertex set $V$. 
There is an infinite family of graphs obtained by adding an arbitrary number $L$ of edges between the vertices of $T$,
since self-loops and multiple edges are allowed. 
Such graphs have $\vert E(G) \vert = \vert V\vert -1 + \vert L\vert $ edges and nullity $L$ (that is $L$ independent loops, 
{since we deal with connected graphs, as already stated above}).
In that case we  write $W(T)= \sum_{G \supset T} w(G,T)$ to indicate summation over the infinite family 
of such $G$'s\footnote{This family could possibly be enlarged or restricted
by additional "Feynman rules" in the context of a QFT with a particular set of edges (propagators) and interactions (vertices), 
but this issue is not important at the level of generality of this paper,
which does not deal with a particular QFT {model}.}.
The corresponding series may of course be divergent or convergent depending on the exact weights considered\footnote{The category of Feynman graphs
to consider for the constructive applications below has in fact slightly more structure, since Feynman graphs have also labeled half-edges, 
according to Wick theorem. It provides them in particular with a canonical ciliated ribbon structure, by labeling the half-edges 
starting from the cilium. These additional (important) subtleties 
are not considered here, as the half-edge labeling will play no role in \emph{this} paper.}.

A complete ordering of the $\vert E(G)\vert $ edges of $G$ is called a {\bf Hepp sector} in QFT terminology \cite{Hepp}.
The set of such orderings, $S(G)$, has $\vert E(G)\vert !$ elements. 
For any such {Hepp} sector $\sigma \in S(G)$, \emph{Kruskal greedy algorithm} defines a particular  tree $T( \sigma)$,
which minimizes $\sum_{\ell \in T}  \sigma (\ell)$ over all  trees of $G$. We call it for short the \emph{leading tree} for $\sigma$.
Let us briefly explain how the algorithm works. 
The algorithm simply picks the first edge $\ell_1$ in $\sigma$ which is not a self-loop. 
The algorithm then picks the next edge $\ell_2$
in $\sigma$ that does not add a cycle to the (disconnected) graph with vertex set $V$ and edge set $\ell_1$ and so on \cite{kruskal}.
Another way to look at it is through a deletion-contraction recursion: following the ordering of the sector $\sigma$, every edge is either deleted
if it is a self-loop or contracted if it is not. The set of contracted edges is exactly the leading tree for $\sigma$.

Remark that this  leading tree $T(\sigma)$ has been considered intensively in the context of perturbative and constructive
renormalization in QFT \cite{Riv}, as it plays an essential role to get sharp 
bounds on renormalized quantities.

Remark also that given any  {Hepp} sector $\sigma $ the (unordered) tree $T(\sigma)$ comes naturally
equipped with an \emph{induced ordering} $\tau$ (the order in which the edges of $T(\sigma)$
are picked by Kruskal's algorithm). The corresponding \emph{ordered} tree will be denoted as $T_\tau$.

\begin{definition} 
A {\bf probability measure} on trees is a set of positive weights $w(G,T)$ for any labeled 
connected graph $G$ and tree $T \subset G$ such that
\bee
\sum_{T \subset G}  w(G,T) =1  \label{bary00} .
\ee 
The measure and the weights $w$ are called {\bf rational} if all $w(G, T) \in {\mathbb Q}$ and 
{they are called} {\bf symmetric} if $ w(G,T) =  w(G^{\nu},T^\nu)$,
where $\nu$ is any permutation of $V$, hence any relabeling of the vertices of $G$ and $T$. 
The measure and the weights are called
{\bf{constructive}} if there exists a $T$-dependent probability measure $(\Omega_T, \Sigma_T, \mu_T)$\footnote{In all 
concrete examples $\Omega_T$
is a topological space, and the sigma-algebra $\Sigma_T$ is its Borel sigma-algebra and will play no further role.}
and a $(T,u)$ dependent real positive symmetric matrix $X^T_{v,v'} (u)$ for any $u \in \Omega_T$, 
with diagonal $X^T_{v,v} (u) =1$ for any $u$, such that 
\bee \label{constructiveweights}
w(G,T) =  \int_{\Omega_T}  d\mu_T (u)  \prod_{\ell \not\in T}   X^T_{i(\ell) j(\ell)} (u)
\ee 
where $v(\ell)$ and $v'(\ell)$ denote, by a slight abuse of notation, the two vertices
that the edge $\ell$ hooks to.
 \end{definition}

Note that the order of the two vertices $v(\ell)$ and $v'(\ell)$ above plays no r\^ole, since the matrix $X$ is symmetric. From a QFT perspective, this comes from the fact that one can endow the internal edges of a Feynman graph with any orientation.

This constructive property is exactly what allows, in the case of stable Bosonic interactions such as $\phi^4$, 
to rewrite any tree amplitude $A_T$ of the loop vertex expansion 
in \eqref{const} as an integral over $\Omega_T$ for the measure $d\mu_T$ of a 
functional integral over a positive Gaussian measure of covariance 
$X^T_{ij} (u)$ of a well-bounded integrand. Hence it is exactly
the property necessary for {inequality} \eqref{const1} to hold. For non-constructive weights, there is no such functional integral
representation.

\bigskip

Let us recall here the definition of the symmetric weights $w_{{s}}(G,T)$ {{(see again \cite{Rivasseau:2013ova})}}:
\begin{definition} The symmetric weights {$w_{{s}}(G,T)$} 
are the percentage of Hepp sectors for which {the tree} $T$ is 
{a} leading {tree}
\bee w_{{s}}(G,T) = \frac{1}{\vert E(G) \vert !}  \sum_{\sigma \in S(G)} \chi( T(\sigma) =T)  \label{def}
\ee
where $\chi(T(\sigma) =T)$ is $1$ if the leading tree for the {Hepp sector} $\sigma$ is {the tree} 
$T$, and $0$ otherwise. 
\end{definition}

Normalization and rationality of these weights are obvious. We then recall the main result of \cite{Rivasseau:2013ova}{{:}
\begin{theorem}
The symmetric weights $w_{{s}}(G,T)$ are constructive.
\end{theorem}
Indeed, {identity} \eqref{constructiveweights} holds for $\Omega_T = [0,1]^{\vert E(T) \vert}$ with Lebesgue measure and 
for the matrix $X^T_{ij} (u)$ which is 1 for $i=j$ and which is for $i \neq j$ the infinimum over the $u_{\ell}$ parameters 
of the lines $\ell$ in $P^T_{ij}$, where $P^T_{ij}$ is the unique path in $T$ between vertices $i$ and $j$  \cite{Rivasseau:2013ova}.
The fact that this matrix is symmetric positive for any $u$ is then a well{-}known property of this infimum function and of the Taylor forest 
formula \cite{BK,AR1,R1}.



\section{Partition {Tree} Weights}

Consider again a fixed vertex set $V$.

A  partition of $V$  into $k $ non empty disjoint subsets
$V = V_1 \cup \cdots {\cup}V_k$ is called {\bf trivial} if $k=1$ and 
{\bf non-trivial} if $k \ge 2$. The subsets
of the partition are also called \textbf{blocks} in what follows. From now on
we suppose we made a choice of a fixed such partition $\Pi$. Our goal is to define,
for any graph $G$ with vertex set $V$
an associated rational constructive measure for the  trees of $G$.

An edge $\ell \in G$ with ends $i$ and $j$ is called {\bf trans-block} for {the partition} $\Pi$ if {the vertices} $i$ and $j$ 
belong to two distinct blocks $V_{k(i)}$ and $V_{k(j)}$ of $\Pi$. Remark that a self-loop is never
trans-block, for any partition.

Given {the graph} $G$ and a trans-block edge $\ell \in G$ with ends $i$ and $j$, we can consider the contracted graph $G/\ell$
in which {the vertices} 
$i$ and $j$ are replaced by a single contracted vertex $\widehat{ij}$ and 
{the edge} $\ell$ is removed. 
This contracted graph is naturally equipped with a {\bf contracted partition}
$\Pi/\ell$ defined by the blocks
$V_1, \cdots {,} V_{k(i)} - \{i\},  \cdots {,} V_{k(j)} - \{j\}, \cdots , V_k, V_{k+1} = \{\widehat{ij}\}$ \emph{any empty block being omitted}. 
Hence it is a partition into $k'$ blocks, with $k-1 \le k' \le k+1$. Remark that this reduced partition 
has always at least a singleton block, namely $V_{k+1}$. Remark also that it can be trivial 
only if the graph $G$ has exactly two vertices; indeed the equation $k'=1$
implies that the initial partition was solely made of $V_{k(i)} = \{i\}$ and $V_{k(j)} = \{j\}$.

Iterating this construction we arrive at the definition
of a trans-block ordered forest:

\begin{definition}
An ordered forest $\cF =\{\ell_1 , \cdots \ell_p\}$ $p \le \vert V \vert -1$ is called {\bf trans-block} for {the partition} $\Pi$ if {the edge} $\ell_1$ is trans-block
for {the partition} $\Pi$, {the edge} $\ell_2$ is trans-block for the contracted graph $G/\ell_1$ and its contracted partition $\Pi/\ell_1$
and so on until the last edge $\ell_p$ of $\cF$  which is trans-block for the contracted graph $G/  \ell_1/\ell_2 {/}\cdots / \ell_{p-1} $ and the 
contracted partition $\Pi/ \ell_1/ \ell_2 /  \cdots / \ell_{p-1} $. \\
The sequence of graphs $\{ G_0 = G, G_1 = G/\ell_1, \cdots{,} G_{p} = G/\ell_1/ \cdots / \ell_{p} \}$ 
and the sequence of partitions $\{\Pi_0 = \Pi, \Pi_1 = \Pi/\ell_1, \cdots{,} \Pi_{p} = \Pi_{{p-1}}/\ell_{p} \}$ is noted $\cS(G, \Pi, \cF)$.
\end{definition}

Remark that the sequence  $\cS(G, \Pi, \cF)$ indeed depends on the ordering of the forest in a critical way.

The maximal trans-block ordered forests are the trans-block ordered trees:
\begin{lemma}
For any ordered trans-block forest $\cF$, the last partition in $\cS(G, \Pi, \cF)$ is trivial (i.e. made of a single block) 
if and only if $\cF$ is a tree, i.e. has exactly $\vert V\vert -1$ edges. 
\end{lemma}
\prf
Consider a trans-block ordered forest $\cF= \{\ell_1, \cdots \ell_{p}   \}$ and the sequence of 
{$p+1$} 
reduced graphs
$G_0 = G, G_1, \cdots G_{p} $.
The number of vertices decreases by exactly one in each step of this sequence, so the last 
graph has a single vertex, hence we reach a trivial partition if and only if $p = \vert V \vert -1$, hence if and only if $\cF$ is a (trans-block) \emph{tree}. 
\qed

\begin{definition} An ordering $\tau$ of a {given tree} $T$ is called {\bf{admissible}} for the partition $\Pi$ if the \emph{ordered} tree $T_\tau$
is trans-block {for the respective partition}. The set of such admissible orderings 
{for a given tree $T$} is {de}noted {by} $\cA^\Pi (T)$.
\end{definition}
The set of admissible orderings is never empty 
{if the respective partition} 
$\Pi$ is non trivial.
Any admissible ordering  $\tau$ of $T$ defines a sequence of contracted graphs and partitions $\cS(G, \Pi, T_\tau)$.
Do not confuse orderings $\tau $ and the Hepp sectors for the {\it full} graph $G$ considered in the previous section. Remark however 
that the orderings $\tau$ can be considered as Hepp sectors for {{\it the tree}} $T$.

The partition weight $w^\Pi(G, T)$ will be defined in {formula} \eqref{defpartweights}  below as {a} sum over all admissible orderings $\tau \in \cA^\Pi(T)$
of certain finite dimensional simple integrals. Their definition requires first that we define {the} so-called {\bf contact indices}.
These indices are defined for any pair of vertices $(v,v')$ of $G$ (including the case $v=v'$) and any ordered trans-block tree $T_\tau$:


\begin{definition}[Contact Indices]
Consider an ordered trans-block tree  $T_\tau= \{\ell_1, \ldots {,}\ell_{\vert V\vert -1} \}$ and its associated 
sequence of reduced graphs and partitions $\cS(G, \Pi, T_\tau) =  \{G_p, \Pi_p\}$  with $0 \le p \le \vert V\vert -1$. 
We define the \textbf{first contact index} $i^{\Pi}_{T_\tau}(v,v')$ as the smallest value of $p$ such that 
{the two vertices} $v$ and $v'$ 
belong to different blocks for $\Pi_p$,
and the \textbf{second contact index}  $j^{\Pi}_{T_\tau}(v,v')$ as the smallest value of $p$
for which $v$ and $v'$ are collapsed into a single reduced vertex in $G_p$.
If $v=v'$
we 
set 
by convention: $i^{\Pi}_{T_\tau}(v,v')  = -1$ and 
$ j^{\Pi}_{T_\tau}(v,v')=0$. 
\end{definition}

 {Let us make the following remark}. If  {the two vertices} $v$ and $v'$ 
belong to distinct blocks of the partition $\Pi${,} 
we have therefore $i^{\Pi}_{T_\tau}(v,v')=0$.

\begin{lemma} The two contact indices obey $i^{\Pi}_{T_\tau}(v,v') < j^{\Pi}_{T_\tau}(v,v')$.
\end{lemma}
\prf {This follows directly from the definition above.}
\qed

\begin{definition}[Contact Matrices]
For any graph with vertex set $V$, any given vertex set partition $\Pi$, any given ordered tree $T_\tau$ and 
$u= \{u_1, \ldots  {,} u_{ \vert V \vert -1} \}$ in $[0,1]^{\vert V \vert -1}$, we define the $\vert V \vert $ 
by $\vert V \vert $ $u$-dependent real symmetric matrix, called the {\bf contact matrix}, 
$X^{\Pi, T_\tau}(u) $ by the following formula:
\bee 
\label{X}
X^{\Pi, T_\tau}_{v,v'}  (u) := \prod_{i^{\Pi}_{T_\tau}(v,v') < k \le j^{\Pi}_{T_\tau}(v,v')}   u_k ,\ \ \ {\forall v,v'\in V}.
\ee
Moreover, we define the {\bf tree edge factor}, by the following formula: 
\bee 
\label{Y}
Y_\ell^{\Pi, T_\tau} (u) :=  \prod_{i^{\Pi}_{T_\tau}(v(\ell),v'(\ell)) < k < j^{\Pi}_{T_\tau}(v(\ell),v'(\ell))  }   u_k  .\ \  \ \
\forall \ell \in T.
\ee
Note that in \eqref{Y} we have again denoted by $v(\ell)$ and $v'(\ell)$ (by the same slight abuse of notation) the two vertices that the (tree) edge $\ell$ hooks to. 
\end{definition}

Remark that $X^{\Pi, {T_\tau}}_{v,v} (\sigma) (u) = 1$, since by convention an empty product is one, hence the matrix $X^{\Pi, T_\tau}(u) $
has diagonal entries all equal to one.

We are now finally in position to define the partition weights associated to an admissible sector $\sigma$.
\begin{definition}[Partition {Tree} Weights]
For  {a set} $V$ and  {a partition} $\Pi$ fixed, the partition weights $w^\Pi(G, T)$  
{associated to any tree $T$ of the graph $G$ (whose vertex set is $V$)} 
are defined as
\bee \label{defpartweights} 
w^\Pi(G, T) := \sum_{\tau \in \cA^\Pi (T)}
\int_	0^1 du_1 \cdots \int_0^1 du_{\vert V \vert -1}   \bigl[    \prod_{\ell \in T}  Y_\ell^{\Pi, T_\tau} (u)  \bigr]  
\bigl[  \prod_{\ell \not\in T}  X^{\Pi , T_\tau}_{v (\ell),v' (\ell)} (u)  \bigr]  
\ee
where $v(\ell)$ and $v'(\ell)$ are the two vertices that the edge $\ell$ hooks to.
\end{definition}

Each  {partition tree weight} $w^\Pi(G, T)$ being obviously a 
sum over $\cA^\Pi (T)$ of positive rational numbers is a positive rational number.
We first prove a lemma stating the normalization of these weights, for a given graph and a given 
partition of its vertex set.

\begin{lemma} \label{barypart}
We have
\bee \sum_{T\subset G}  w^\Pi(G,T) =1 . \label{barypart1}
\ee
\end{lemma}
\noindent{\bf Proof:} 
The key idea is to Taylor expand  the function $1= [\prod_{\ell \in G} x_\ell ]\vert_{x_\ell =1 \, \forall \ell} $ 
according to a $\Pi$-dependent recursion with $\vert V \vert -1$ steps which, at each step $i$, uses
an elementary first order Taylor expansion with integral remainder 
\beqa
\label{baza}
f(1) = f(0) + \int_0^1 f'(u_i) du_i, \quad i=1,\ldots , \vert V \vert -1. 
\eeqa
This recursion ``builds''
step by step a sum over all trans-block ordered trees.


\medskip

Let  $F_0 (x_1, \ldots, x_{|E|}):=x_1\ldots x_{|E|}.$ 
The first step of the induction requires first to multiply each variable $x$ corresponding to a trans-block edge for the partition 
$\Pi_0$ with a dummy variable $u_1$. This amounts to consider the function
\beqa
F_0 (u_1 x_1,\ldots,u_1 x_{k_0}, x_{k_0+1}, \ldots, x_{|E|})\vert_{x_i=1, \forall i},
\eeqa
where, without any loss of generality we have placed on the first $k_0$ positions the $k_0$ variables associated to the trans-block edges for the partition $\Pi_0$. Note that 
since $\Pi_0$ is non-trivial and $G$ is connected, one has $k_0 > 0$. Using the 
multi-variable version of the Taylor expansion \eqref{baza} leads to
\beqa
\label{i1}
1&=&F(0,\ldots,0, 1,\ldots,1)\\
&+& \sum_{\ell_1}\int_0^1 d u_1 \frac{\partial}{\partial x_{\ell_1}}
F_0 (u_1 x_1,\ldots,u_1 x_{k_0}, x_{k_0+1}, \ldots, x_{|E|})\vert_{x_i=1, \forall i}, \nonumber
\eeqa
where the sum over $\ell_1$ runs over the $k_0$ trans-block edges for the partition $\Pi_0$. Note that the 
first term on the RHS of equation \eqref{i1} vanishes (because of the definition of the function 
$F_0$ and because $k_0\ne 0$, as already noticed above).
For the sake of simplicity, let us fix the trans-block edge in the sum \eqref{i1}. It will be the first edge in the recursive construction of our ordered tree. 
In the RHS of equation \eqref{i1} the derivative 
leads to 
\beqa
\label{i2}
1=    \sum_{\ell_1}\int_0^1 d u_1 u_1^{k_0 -1} F_1.
\eeqa
In the function $F_1$ 
we fix definitely $x_{\ell_1} =1$, so that $F_1$ no longer depends on it.
We can now proceed with the next induction, which is identical but for the graph $G_1 = G/\ell_1$ and the partition $\Pi_1 = \Pi /\ell_1$. 

At a generic step $n$, this expansion leads to
\beqa
\label{in}
1=  \sum_{\ell_1, \ldots,  \ell_n}   \int_0^1 d u_1 \ldots d u_n u_1^{k_0 - 1} 
\ldots u_n^{k_{n-1} - 1}
F_n 
\eeqa
where the exponent $k_{j}$ represents the number of trans-block edges for the partition $\Pi_j$ in our sequence of partitions.  As above, the factor $F_n$ does not depend on the variables associated to the contracted edges $\ell_1, \ldots, \ell_n$. This process continues until there are no more trans-block edges for the respective partition; this happens at the $(|V|-1)$th step. The corresponding factor $F_{|V|-1}$ is then no longer interpolated hence equal to $1$.
The expansion therefore results in a sum over all trees and all their admissible sectors, renaming dummy variables
$u_{}$  
\beqa
\label{ifinal}
1=   \sum_{T, \tau \in A^\Pi (T) ; \; T_{\tau} = \ell_1, \ldots, \ell_{\vert V \vert -1}  }   \int_0^1 d u_1 \ldots d u_{|V|-1} u_1^{k_{0}- 1} 
\ldots u_{|V|-1}^{k_{|V|-2}  -1} .
\eeqa

Let us now identify, for a fixed ordered tree $T_\tau $ in this sum, the integrand $u_1^{k_{0}- 1} \ldots u_{|V|-1}^{k_{|V|-2}  -1}$ with the 
two products appearing in \eqref{defpartweights}:
\beqa
\label{factor} \prod_{\ell \in T} Y_\ell^{\Pi, T_\tau}\prod_{\ell\notin T} X^{\Pi, T_\tau}_{v(\ell),v'(\ell)}  \; = \; u_1^{k_{0}- 1} \ldots u_{|V|-1}^{k_{|V|-2}  -1}.
\eeqa
Indeed within the LHS of \eqref{factor} - which is, by construction, a product of various powers of the (tree) edge variables $u_k$ ($k=1,\ldots,|V|-1$) - the variable $u_1$ appears as many times as there are edges $\ell$ with contact indexes $i=0$ and $j\ge 1$. This is nothing but the number of edges which are trans-block for the partition $\Pi_0$ minus $1$, the minus 1 correction coming from the fact that, for $\ell_1$, the contact indices are $0$ and $1$, and the factor $u_1$ is not taken into account, by definition, within the tree factor $Y_{\ell_1}^{\Pi, T_\tau}$. 
The analogous reasoning holds for $u_k$ ($k=2,\ldots,|V|-1$) and this concludes the proof. \qed



\bigskip

Let us now rewrite formula  \eqref{defpartweights} as
\beqa \label{defpartweights2} 
w^\Pi(G, T) = \sum_{\tau \in \cA^\Pi (T)} w^\Pi(G, T^\tau) ,
\eeqa
where we define:
\beqa
\label{poids22}
w^\Pi(G, T^\tau) :=
\int_	0^1 du_1 \cdots \int_0^1 du_{\vert V \vert -1}   \bigl[    \prod_{\ell \in T}  Y_\ell^{\Pi, T_\tau} (u)  \bigr]  
\bigl[  \prod_{\ell \not\in T}  X^{\Pi , T_\tau}_{v (\ell),v' (\ell)} (u)  \bigr]. 
\eeqa

One has:

\begin{corollary}
For a given graph $G$, vertex partition $\Pi$ and ordered tree $T^\tau$, 
the weights defined in \eqref{poids22} can be written as
\beqa \label{defpart}
w^\Pi(G, T^\tau)=\prod_{i=0}^{\vert V\vert -2} \frac{1}{k_i},
\eeqa 
where $k_i$ give the number of trans-block edges for the partition $\Pi_i$ in the corresponding partition sequence.
\end{corollary}
\prf This is a direct consequence of equation \eqref{in}.
\qed

\medskip

Let us mention here that a somehow similar use of the Taylor formula with 
integral reminder  was made in \cite{malek}, within the cluster expansion framework.
The use of the integral reminder is one possible way of not dealing with the whole perturbative series, since this series most often diverges in QFT.


\medskip

Let us now prove the following lemma:

\begin{lemma} \label{positpart}
The symmetric matrix $ X^{\Pi, T_\tau}_{v,v'} (u)$ is positive semi-definite.
\end{lemma}
\noindent{\bf Proof:}
We prove the positivity for any fixed ordered tree $T_\tau$ again by a recursion based upon the sequence of graphs
$G_0, \ldots, G_{\vert V\vert -1}$ and the associated partitions $\Pi_0,\ldots, \Pi_{\vert V\vert -1}$.

We define first for any $n$ by $n$ matrix $X$ and any partition $\Pi$ of $[1, \ldots , n]$, the 
projected matrix $X^\Pi$ which has elements $X^\Pi_{ij} = X_{ij}$ if $i$ and $j$ belong to the same block
of $\Pi$ and 0 otherwise. We then remark that for any symmetric positive $X$, $X^\Pi$ is positive and 
for any real $u \in [0,1]$ the interpolated matrix 
$X (u) = u X + (1-u )X^\Pi$ is also positive, as barycentric combination of two positive matrices.

Let us now define the positive $\vert V \vert$ by $\vert V \vert $ matrix $X_0$, which has matrix elements all equal to one.

We then define:
\bea
X_1 (u_1) &=& u_1 X_0 + (1- u_1) X_0^{\Pi_0} \nonumber\\
X_2 (u_1, u_2) &=& u_2 X_1 (u_1) + (1- u_2) X_1^{\Pi_1} (u_1)  \nonumber \\
& \dots & \nonumber \\
X_{\vert V \vert -1} ( u_1 , \ldots  , u_{\vert V \vert -1} )  &=&   u_{\vert V \vert -1} X_{\vert V \vert -2} (u_1, \ldots, u_{\vert V \vert -2}) 
\nonumber \\&+& (1- u_{\vert V \vert -1}) Xu_{\vert V \vert -2}^{\Pi_{\vert V \vert -2}} (u_1, \ldots , u_{\vert V \vert -2}) .
\eea
Note that all these matrices are positive, again as barycentric combinations of two positive matrices.
This mechanism thus leads that to a ``final'' $\vert V \vert$ by $\vert V \vert $ matrix $X_{\vert V \vert -1}$ for which 
the matrix element corresponding to the entry $(v,v')$ ($v,v'=1,\ldots, \vert V \vert$) 
is a product of variables $u_j$ ($j=1,\ldots, \vert V \vert -1$). 
Each such variable $u_j$ is present in this product if and only if 
the edge (or the edges, since multi-edges are allowed) connecting the vertices $v$ and $v'$ in $G_{j-1}$ 
is a trans-block edge for 
the corresponding partition  $\Pi_{j-1}$.

We have thus ``constructed'' the matrix given by the formula \eqref{X}, since the combinatorial definition of the 
contact indices lead to the same product
of the tree variables $u_j$. This concludes the proof.
\qed

The main result of this paper is the following theorem:
\begin{theorem} \label{theopart}
For any {set} $V$ and {partition} $\Pi$ the partition weights $w^\Pi(G, T)$ 
on the trees $T$ of any graph $G$ with vertex set $V$
 form a rational constructive probability measure.
\end{theorem}
\prf 
The claim follows from Lemmas \ref{barypart} and \ref{positpart}. Indeed the set $\Omega_T$
is the disjoint union for all admissible $\tau$'s of a distinct copy of $[0, 1]^{\vert V \vert -1} $ with measure 
\bee d\mu_{T_\tau} =  \prod_{\ell \in T} Y^{\Pi , T_\tau}_\ell (u)\;  \prod_{k=1}^{\vert V \vert -1} du_k .
\ee
Remarking that the normalization of the full measure $d \mu_T$ is nothing but equation \eqref{barypart1} for the particular case
when $G=T$ concludes  the proof. \qed


\bigskip

Before ending this section, let us mention that an important particular case of our results can be obtained when the vertex partition is made of a singleton plus a single block
containing all the remaining vertices.
As already mentioned in the introduction, the mechanism exposed in this paper leads to constructive tree weights for a {\it rooted graph} (the singleton being the root of the graph).
An example of such tree weights for a rooted graph is given in subsection 
\ref{ex-rooted} below.
This is then generalized for {\it multi-rooted graphs}, which correspond to vertex partitions made of several singletons and a remaining block containing the rest of the vertices. An example of a double-rooted graph is analyzed in detail in subsection \ref{2singleton} below.

\section{Examples}

\subsection{Symmetric weights - complete partition}

Symmetric weights correspond to the symmetric partition $\Pi^s$ of $V$ into 
$\vert V \vert$ singletons. Let us check directly that this is indeed the case, namely 
that $w_s(G,T)$, as defined in \eqref{def} is equal to 
$w^{\Pi^s}(G,T)$ defined by formula \eqref{defpart}:
\begin{lemma} The symmetric weights $w_s (G,T)$ 
are the partition weights for the partition $\Pi^s$ of $V$ into 
$\vert V \vert$ singletons: 
\bee \label{singletonseq} w_s(G,T) = \sum_{\sigma  \vert T(\sigma) =T }  1/\vert E(G)\vert !  =  \sum_\tau  \prod_{i=0}^{\vert V\vert -2} \frac{1}{k_i} = w^{\Pi^s} (G,T),
\ee 
where the sum over $\tau $ is performed over all the orderings of $T$ and
$k_i$ is the number of trans-block edges for the partition $\Pi^s_i$ in the partition sequence corresponding to $T_\tau$,
starting from the all-singletons partition $\Pi^s$.
\end{lemma}
\prf  Remark that in the symmetric case every sector is admissible, hence there is no restriction on the sum over $\tau$. 
We work by induction on the number of vertices of $G$ in \eqref{singletonseq}. 
Let us start with an initial general graph $G=G_0$, and suppose it has 
a certain set $L_0$ of tadpole edges. Consider the graph $G'_0= G_0 - L_0$ with all tadpoles of $G$ deleted. 
Since the weights $w_s (G,T)$ cannot depend on the position of the tadpoles edges in the Hepp sector $\sigma$, we have
$w_s(G_0,T) = w_s(G'_0,T)$. 
In $G'_0$, which has no tadpoles, all edges are trans-block at first step
(since $\Pi_0 = \Pi^s$ is made of singletons). Therefore the factor $k_0$ in \eqref{defpart} is $k_0 = \vert E(G'_0)\vert  $. We can write 
\bea w_s(G,T) &=& \sum_{\ell_1 = \sigma (1) \in T} \frac{1}{k_0}    \sum_{\sigma_1 \; \vert\; T(\sigma_1) = T - \ell_1}  \frac{1}{(k_0-1)!}  \nonumber\\
&=& \sum_{\ell_1 = \sigma (1) \in T}  \frac{1}{k_0}  w_s(G_1, T_1) \nonumber\\
&=& \sum_{\ell_1 = \sigma (1) \in T}  \frac{1}{k_0}  w^{\Pi^s}(G_1, T_1) \label{singconclu}
= w^{\Pi^s}(G, T)
\; .\eea 
where $\ell_1$ is the first edge of $T_\tau$;
$G_1$ is obtained by contracting the edge $\ell_1$ in $G'_0$, $T_1$ is obtained by contracting
the first edge $\ell_1$ in $T_\tau$ and the sum over $\sigma_1$ runs over the Hepp sector of $G_1$.
Since $G_1$ has one vertex less than $G_0$ we used the induction hypothesis in the last line of \eqref{singconclu} to conclude. \qed

\medskip

Consider the particular example of the graph of Fig. 2. The symmetric weights are:
\bea  &&w_s(G,T_{125})= w_s(G,T_{126})= w_s(G,T_{156}) =w_s(G,T_{256})= 1/15, \nonumber \\
 &&w_s(G,T_{135})= w_s(G,T_{136})= w_s(G,T_{235}) =w_s(G,T_{236}) = w_s(G,T_{145})\nonumber \\ 
 && \quad\quad\quad\quad\quad =  w_s(G,T_{146}) = w_s(G,T_{245})= w_s(G,T_{246}) = 11/120  . 
\label{symexample}
\eea 
These weights were computed in \cite{Rivasseau:2013ova} (note the different labeling we use here with respect to the one of \cite{Rivasseau:2013ova}).

\subsection{One singleton partition - rooted graph}
\label{ex-rooted}

The next case we deal with is the one when the partition is made of a certain number $p$ 
of singletons plus a single block with all other remaining vertices. 
As already mentioned above, when $p=1$, the partition corresponds to 
work on a rooted graph.
We obtain weights related to the Brydges-Battle-Federbush
constructive QFT approach (see \cite{BF1,BF2,BF3}).
When the number $p$ of singletons is at least two, the weights correspond to multi-rooted weights (see the example in subsection \ref{2singleton}.

Consider the graph of Figure 1.
\begin{figure}[!htb] \label{exa}
\centering
\includegraphics[scale=0.8]{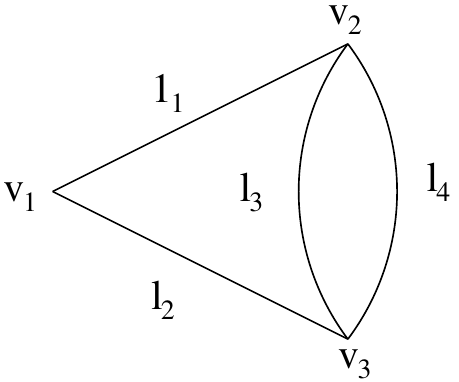}
\caption{An example of a three vertex graph.}
\end{figure} 
Let us find the tree weights in the case of root at $v_1$. This correspond to the partition
\beqa
\Pi_1=[\{v_1\},\{v_2,v_3\}].
\eeqa
There are six admissible ordered trees, namely   $T_{12}$, $T_{13}$, $T_{14}$, $T_{21}$, $T_{23}$ and finally $T_{24}$.
The weights are 
\bee  
w^{\Pi_1}(G,T_{12})  = w^{\Pi_1}(G,T_{21})  =  \int_	0^1 du_1 du_2   u_1 . (u_2)^2  =  1/6   
\ee
since   $i(l_1)= 0$, $j(l_1)= 1$; $i(l_2)= 0$, $j(l_2)= 2$; $i(l_3)= 1$, $j(l_3)= 2$; $i(l_4)= 1$, $j(l_4)= 2$, where, from now on, we have simplify the notations for the contact indices (since an edge is identified with a pair of (non-necessarily distinct) vertices of the graph).

Similarly, one has
\bea w^{\Pi_1}(G,T_{13})  &=&  w^{\Pi_1}(G,T_{14})  =   w^{\Pi_1}(G,T_{23})  = w^{\Pi_1}(G,T_{24})  \nonumber
\\ &=& \int_0^1 du_1 du_2   (u_1 u_2) . (u_2)  =  1/6    
\eea
since  $i(l_1)= 0$, $j(l_1)= 1$; $i(l_2)= 0$, $j(l_2)= 2$; $i(l_3)= 1$, $j(l_3)= 2$; $i(l_4)= 1$, $j(l_4)= 2$.

We can check that $\sum_{T\subset G}  w^{\Pi_1}(G,T) =1$.

\bigskip

Let us now count the factors in the case of the root at $v_2$. This corresponds, in the formalism of this paper, to consider the partition:
\beqa
\Pi_2=[\{v_2\},\{v_1,v_3\}].
\eeqa
As above, there are seven admissible ordered trees, namely   $T_{12}$, $T_{13}$, $T_{14}$, $T_{31}$, $T_{32}$,  $T_{41}$ and $T_{42}$.
The associated weights compute to: 
\bee  w^{\Pi_2}(G,T_{12})  =  \int_0^1 du_1 du_2  [ 1 . 1 .  (u_1 u_2) .(u_1 u_2)  ]  =  1/9   
\ee
since  $i(l_1)= 0$, $j(l_1)= 1$; $i(l_2)= 1$, $j(l_2)= 2$; $i(l_3)= 0$, $j(l_3)= 2$; $i(l_4)= 0$, $j(l_4)= 2$,
and 
\bee  w^{\Pi_2}(G,T_{13})  = w^{\Pi_2}(G,T_{14})   = \int_0^1 du_1 du_2  [1 . u_2 .  u_1 .  (u_1 u_2) ] =  1/9    
\ee
since  $i(l_1)= 0$, $j(l_1)= 1$; $i(l_2)= 1$, $j(l_2)= 2$; $i(l_3)= 0$, $j(l_3)= 2$; $i(l_4)= 0$, $j(l_4)= 2$.
and
\bee 
w^{\Pi_2}(G,T_{31})  =   w^{\Pi_2}(G,T_{41})   = \int_0^1 du_1 du_2  u_1 .  u_2 . 1. u_1 =  1/6   
\ee
since $i(l_1, T_{31})= 0$, $j(l_1, T_{31})= 2$; $i(l_2, T_{31})= 1$, $j(l_2, T_{31})= 2$; $i(l_3, T_{31})= 0$, $j(l_3, T_{31})= 1$; $i(l_4, T_{31})= 0$, $j(l_4, T_{31})= 1$.
and
\bee 
w^{\Pi_2}(G,T_{32})  = w^{\Pi_2}(G,T_{42})   = \int_0^1 du_1  du_2  u_1u_2  . 1. 1.  u_1   = 1/6    
\ee
since
$i(l_1, T_{32})= 0$, $j(l_1, T_{32})= 2$; $i(l_2, T_{32})= 1$, $j(l_2, T_{32})= 2$; $i(l_3, T_{32})= 0$, $j(l_3, T_{32})= 1$; $i(l_4, T_{32})= 0$, $j(l_4, T_{32})= 1$.
As expected, we have again: $\sum_{T \subset G}  w^{\Pi_2}(G,T) =1$.

\subsection{Two singleton partition - multi-rooted graph}
\label{2singleton}


We consider the graph of Fig. 2 with four vertices and six edges, for the partition 
\beqa
\Pi =[ \{v_1\}; \{v_2\}; \{v_3, v_4\} ].
\eeqa
This corresponds to considering a graph with {\it two roots}, the first at the vertex $v_1$ and the second at the vertex $v_2$.
\begin{figure}[!htb] \label{graf22}
\centering
\includegraphics[scale=0.8]{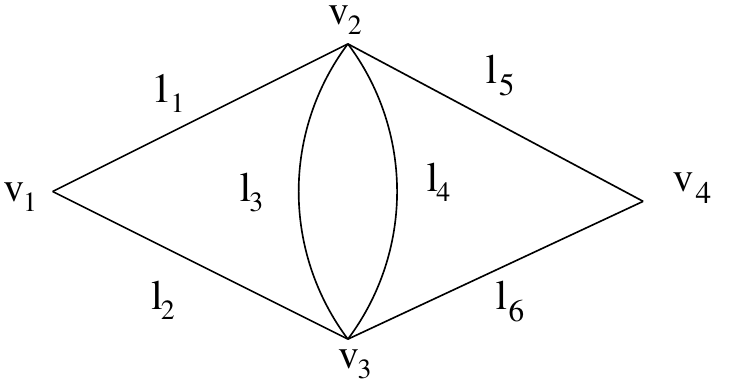}
\caption{An example of a four vertex graph.}
\end{figure} 
This graph has twelve trees:
\beqa
\label{list-trees}
&& T_{125}, T_{135}, T_{145}, T_{235}, T_{245},
\nonumber\\
&& T_{236}, T_{246}, T_{256},
\nonumber\\
&& T_{126}, T_{136}, T_{146}, T_{156}.
\eeqa
The first five trees in this list (the trees in the first line of \eqref{list-trees} above) can be each endowed with 
six admissible orders (each order is for these trees admissible). The next three trees in the list 
(the trees in the second line of \eqref{list-trees}) can be endowed with only four admissible orders,
while the last four trees (the trees in the last line of \eqref{list-trees}) can be endowed with 
three admissible orders. This makes up for a total of fifty-four ordered trans-block trees to consider for this graph.

Let us explictly consider the first of these ordered trees, namely the  $(l_1, l_2,l_5)$ ordered tree.
For the three tree lines, the contact indices are: $i(l_1)=0, j(l_1)=1$, 
 $i(l_2)=0, j(l_2)=2$ and $i(l_5)=0, j(l_5)=3$. For the remaining three loop lines, the contact indices are:
 $i(l_3)=0, j(l_3)=2$,  $i(l_4)=0, j(l_4)=2$  and finally $i(l_6)=0, j(l_6)=1$. This leads to the contribution:
\beqa
\label{54}
\int_0^1 d u_1 du_2 du_3 u_1^4 u_2^3u_3=\frac{1}{40}.
\eeqa
The other five admissible orders for this tree lead to the weights $1/80$, $1/50$, 
$1/100$, $1/100$ and finally, again $1/100$. Thus, for the 
total of six admissible orders that one can endow this tree with, one obtains a total weight of $7/80$.

\medskip

After a tedious but straightforward computation, we obtain all forty-eight admissible order contributions 
and find the complete list of all tree weights for this partition:
\bea  
&&w^\Pi (G,T_{135}) = w^\Pi(G,T_{145})= 47/400 , \nonumber \\ && w^\Pi(G,T_{235})  = w^\Pi(G,T_{245})=  11/100, \nonumber 
\\ && w^\Pi(G,T_{236}) =  w^\Pi(G,T_{246}) = 2/25,\nonumber  \\ &&
  w^\Pi(G,T_{136}) = w^\Pi(G,T_{146}) = 3/40,\nonumber 
  \\ &&
w^\Pi(G,T_{256}) = 1/20, \quad w^\Pi(G,T_{126})  = 11/200, \nonumber 
\\ && w^\Pi(G,T_{125})= 7/80, \quad w^\Pi(G,T_{156}) = 17/400.
\eea 
Note that these tree weights are different from the symmetric weights of 
\eqref{symexample}.
Finally, one can check that $\sum_{T\in G} w^\Pi (G, T)=1$, as expected.

\medskip
\noindent{\bf Acknowledgments}
A. Tanasa  is partially supported by the "Combinatoire  alg\'ebrique" Univ. Paris 13, Sorbonne Paris Cit\'e BQR  grant, "Cartes 3D" CNRS PEPS grant as well as
the grants PN 09 37 01 02 and CNCSIS Tinere Echipe 77/04.08.2010. Research at Perimeter Institute is supported by the Government of Canada through Industry Canada and by the Province of
Ontario through the Ministry of Research and Innovation.

\vspace{2cm}

\noindent
{\small ${}^{a}${\it LPT, 
CNRS UMR 8627, Univ. Paris 11, 91405 Orsay Cedex, France, EU}}\\
{\small ${}^{b}${\it Perimeter Institute for Theoretical Physics, 31 Caroline St. N, ON, N2L 2Y5, 
Waterloo, Canada}}\\
{\small ${}^{c}${\it Universit\'e Paris 13, Sorbonne Paris Cit\'e, 99, avenue Jean-Baptiste Cl\'ement \\
LIPN, Institut Galil\'ee, 
CNRS UMR 7030, F-93430, Villetaneuse, France, EU}}\\
{\small ${}^{d}${\it 
Horia Hulubei National Institute for Physics and Nuclear Engineering,\\
P.O.B. MG-6, 077125 Magurele, Romania, EU}}\\

\end{document}